\documentstyle[12pt]{article}

\skewchar\fivmi='177
\skewchar\sixmi='177
\skewchar\sevmi='177
\skewchar\egtmi='177
\skewchar\ninmi='177
\skewchar\tenmi='177
\skewchar\elvmi='177
\skewchar\twlmi='177
\skewchar\frtnmi='177
\skewchar\svtnmi='177
\skewchar\twtymi='177
\def\@magscale#1{ scaled \magstep #1}
\font\twfvmi  = ammi10   \@magscale5 
\skewchar\twfvmi='177
\skewchar\fivsy='60
\skewchar\sixsy='60
\skewchar\sevsy='60
\skewchar\egtsy='60
\skewchar\ninsy='60
\skewchar\tensy='60
\skewchar\elvsy='60
\skewchar\twlsy='60
\skewchar\frtnsy='60
\skewchar\svtnsy='60
\skewchar\twtysy='60
\font\twfvsy  = amsy10   \@magscale5 
\skewchar\twfvsy='60

\catcode`@=11
\def\un#1{\relax\ifmmode\@@underline#1\else
        $\@@underline{\hbox{#1}}$\relax\fi}
\catcode`@=12

\let\du=\d                      
\let\um=\H                      

\def\a{\alpha}
\def\b{\beta}

\def\d{\delta}
\def\e{\epsilon}

\def\g{\gamma}

\def\l{\lambda}
\def\m{\mu}

\def\s{\sigma}

\def\z{\zeta}

\def\L{\Lambda}

\def\S{\Sigma}

\font\sc=font005                        
\def\Sc#1{{\hbox{\sc #1}}}      
\font\ooo=circle10                      
\font\ro=manfnt                         
\def\kcl{{\hbox{\ro 6}}}                
\def\kcr{{\hbox{\ro 7}}}                
\def\ktl{{\hbox{\ro \char'134}}}        
\def\ktr{{\hbox{\ro \char'135}}}        
\def\kbl{{\hbox{\ro \char'136}}}        
\def\kbr{{\hbox{\ro \char'137}}}        

\def\ip{{=\!\!\! \mid}}                                    
\def\bo{{\raise.15ex\hbox{\large$\Box$}}}               
\def\pr{\prod}                                          
\def\TH{{\raise.2ex\hbox{$\displaystyle \bigodot$}\mskip-4.7mu \llap H \;}}
\def\face{{\raise.2ex\hbox{$\displaystyle \bigodot$}\mskip-2.2mu \llap {$\ddot
        \smile$}}}                                      

\def\sp#1{{}^{#1}}                              
\def\Tilde#1{\widetilde{#1}}                    
\def\Hat#1{\widehat{#1}}                        
\def\Bar#1{\overline{#1}}                       
\def\leftrightarrowfill{$\mathsurround=0pt \mathord\leftarrow \mkern-6mu
        \cleaders\hbox{$\mkern-2mu \mathord- \mkern-2mu$}\hfill
        \mkern-6mu \mathord\rightarrow$}
\def\dvec#1{\vbox{\ialign{##\crcr
        \leftrightarrowfill\crcr\noalign{\kern-1pt\nointerlineskip}
        $\hfil\displaystyle{#1}\hfil$\crcr}}}           
\def\dt#1{{\buildrel {\hbox{\LARGE .}} \over {#1}}}     

\def\frac#1#2{{\textstyle{#1\over\vphantom2\smash{\raise.20ex
        \hbox{$\scriptstyle{#2}$}}}}}                   
\def\ha{\frac12}                                        
\def\sfrac#1#2{{\vphantom1\smash{\lower.5ex\hbox{\small$#1$}}\over
        \vphantom1\smash{\raise.4ex\hbox{\small$#2$}}}} 
\def\bfrac#1#2{{\vphantom1\smash{\lower.5ex\hbox{$#1$}}\over
        \vphantom1\smash{\raise.3ex\hbox{$#2$}}}}       
\def\afrac#1#2{{\vphantom1\smash{\lower.5ex\hbox{$#1$}}\over#2}}    

\newskip\humongous \humongous=0pt plus 1000pt minus 1000pt
\def\caja{\mathsurround=0pt}
\def\eqalign#1{\,\vcenter{\openup2\jot \caja
        \ialign{\strut \hfil$\displaystyle{##}$&$
        \displaystyle{{}##}$\hfil\crcr#1\crcr}}\,}
\newif\ifdtup
\def\panorama{\global\dtuptrue \openup2\jot \caja
        \everycr{\noalign{\ifdtup \global\dtupfalse
        \vskip-\lineskiplimit \vskip\normallineskiplimit
        \else \penalty\interdisplaylinepenalty \fi}}}
\def\li#1{\panorama \tabskip=\humongous                         
        \halign to\displaywidth{\hfil$\displaystyle{##}$
        \tabskip=0pt&$\displaystyle{{}##}$\hfil
        \tabskip=\humongous&\llap{$##$}\tabskip=0pt
        \crcr#1\crcr}}

\def\ref#1{$\sp{#1)}$}

\parskip=\medskipamount                 
\thicklines                         

\def\oldheadpic{                                
        \setlength{\unitlength}{.4mm}
        \thinlines
        \par
        \begin{picture}(349,16)
        \put(325,16){\line(1,0){4}}
        \put(330,16){\line(1,0){4}}
        \put(340,16){\line(1,0){4}}
        \put(335,0){\line(1,0){4}}
        \put(340,0){\line(1,0){4}}
        \put(345,0){\line(1,0){4}}
        \put(329,0){\line(0,1){16}}
        \put(330,0){\line(0,1){16}}
        \put(339,0){\line(0,1){16}}
        \put(340,0){\line(0,1){16}}
        \put(344,0){\line(0,1){16}}
        \put(345,0){\line(0,1){16}}
        \put(329,16){\oval(8,32)[bl]}
        \put(330,16){\oval(8,32)[br]}
        \put(339,0){\oval(8,32)[tl]}
        \put(345,0){\oval(8,32)[tr]}
        \end{picture}
        \par
        \thicklines
        \vskip.2in}
\def\oldtitle#1#2#3#4{\oldheadpic\begin{center}\vglue.5in{\large\bf #1}\\[.6in]
        {#2}\\[.1in] {\it Department of Physics and Astronomy}\\
        {\it University of Maryland, College Park, MD 20742}\\[.6in]
        Physics Publication \#{#3}\\ {#4}\\[1.5in] {\bf ABSTRACT}\\[.1in]
        \end{center} \begin{quotation}}                 
\def\oldTitle#1#2#3#4#5#6#7{\oldheadpic\begin{center} \vglue .4in
        {\large\bf #1}\\[.4in]
        {#2}\\[.1in] {\it Department of Physics and Astronomy}\\
        {\it University of Maryland, College Park, MD 20742}\\[.1in]
        {#3}\\[.1in] {\it {#4}}\\ {\it {#5}}\\[.4in]
        Physics Publication \#{#6}\\ {#7}\\[.5in] {\bf ABSTRACT}\\[.1in]
        \end{center} \begin{quotation}}                 
\def\border{                                            
        \setlength{\unitlength}{1mm}
        \newcount\xco
        \newcount\yco
        \xco=-24
        \yco=12
        \begin{picture}(140,0)
        \put(\xco,\yco){$\ktl$}
        \advance\yco by-1
        {\loop
        \put(\xco,\yco){$\kcl$}
        \advance\yco by-2
        \ifnum\yco>-240
        \repeat
        \put(\xco,\yco){$\kbl$}}
        \xco=158
        \yco=12
        \put(\xco,\yco){$\ktr$}
        \advance\yco by-1
        {\loop
        \put(\xco,\yco){$\kcr$}
        \advance\yco by-2
        \ifnum\yco>-240
        \repeat
        \put(\xco,\yco){$\kbr$}}
        \put(-20,11){\tiny University of Maryland Elementary Particle
Physics University of Maryland Elementary Particle Physics University of
Maryland Elementary Particle Physics}
        \put(-20,-241.5){\tiny University of Maryland Elementary
Particle Physics University of Maryland Elementary Particle Physics
University of Maryland Elementary Particle Physics}
        \end{picture}
        \par\vskip-8mm}
\def\bordero{                                           
        \setlength{\unitlength}{1mm}
        \newcount\xco
        \newcount\yco
        \xco=-24
        \yco=12
        \begin{picture}(140,0)
        \put(\xco,\yco){$\ktl$}
        \advance\yco by-1
        {\loop
        \put(\xco,\yco){$\kcl$}
        \advance\yco by-2
        \ifnum\yco>-240
        \repeat
        \put(\xco,\yco){$\kbl$}}
        \xco=158
        \yco=12
        \put(\xco,\yco){$\ktr$}
        \advance\yco by-1
        {\loop
        \put(\xco,\yco){$\kcr$}
        \advance\yco by-2
        \ifnum\yco>-240
        \repeat
        \put(\xco,\yco){$\kbr$}}
        \put(-20,12)
{\ooo
bacdefghidfghghdhededbihdgdfdfhhdheidhdhebaaahjhhdahbahgdedgehgfdiehhgdigicba}
        \put(-20,-241.5)
{\ooo
ababaighefdbfghgeahgdfgafagihdidihiidhiagfedhadbfdecdcdfagdcbhaddhbgfchbgfd
acfediacbabab}
        \end{picture}
        \par\vskip-8mm}
\def\headpic{                                           
        \indent
        \setlength{\unitlength}{.4mm}
        \thinlines
        \par
        \begin{picture}(29,16)
        \put(165,16){\line(1,0){4}}
        \put(170,16){\line(1,0){4}}
        \put(180,16){\line(1,0){4}}
        \put(175,0){\line(1,0){4}}
        \put(180,0){\line(1,0){4}}
        \put(185,0){\line(1,0){4}}
        \put(169,0){\line(0,1){16}}
        \put(170,0){\line(0,1){16}}
        \put(179,0){\line(0,1){16}}
        \put(180,0){\line(0,1){16}}
        \put(184,0){\line(0,1){16}}
        \put(185,0){\line(0,1){16}}
        \put(169,16){\oval(8,32)[bl]}
        \put(170,16){\oval(8,32)[br]}
        \put(179,0){\oval(8,32)[tl]}
        \put(185,0){\oval(8,32)[tr]}
        \end{picture}
        \par\vskip-6.5mm
        \thicklines}
\def\title#1#2#3#4{\border\headpic {\hbox to\hsize{#4 \hfill UMDEPP #3}}\par
        \begin{center} \vglue .5in {\large\bf #1}\\[.6in]
        {#2}\\[.1in] {\it Department of Physics and Astronomy}\\
        {\it University of Maryland, College Park, MD 20742}\\[1.5in]
        {\bf ABSTRACT}\\[.1in] \end{center} \begin{quotation}}  
\def\Title#1#2#3#4#5#6#7{\border\headpic
        {\hbox to\hsize{#7 \hfill UMDEPP #6}}\par
        \begin{center} \vglue .4in {\large\bf #1}\\[.4in]
        {#2}\\[.1in] {\it Department of Physics and Astronomy}\\
        {\it University of Maryland, College Park, MD 20742}\\[.1in]
        {#3}\\[.1in] {\it {#4}}\\ {\it {#5}}\\[.5in] {\bf ABSTRACT}\\[.1in]
        \end{center} \begin{quotation}}                 
\def\endtitle{\end{quotation}\newpage}                  

\def\sect#1{\bigskip\medskip \goodbreak \noindent{\bf {#1}} \nobreak \medskip}
\def\refs{\sect{REFERENCES} \footnotesize \frenchspacing \parskip=0pt}
\def\Item{\par\hang\textindent}

\begin{document}


\def\gg{{\hbox{\sc g}}}
\def\nt{$~N=2$~}
\def\gg{{\hbox{\sc g}}}
\def\nt{$~N=2$~}
\def\tr{{\rm tr}}
\def\Tr{{\rm Tr}}
\def\mpl#1#2#3{Mod.~Phys.~Lett.~{\bf A{#1}} (19{#2}) #3}

\def\scst{\scriptstyle}
\def\itrema{$\ddot{\scriptstyle 1}$}
\def\Bo{\bo{\hskip 0.03in}}
\def\lrad#1{ \left( A {\buildrel\leftrightarrow\over D}_{#1} B\right) }

\def\ula{{\underline a}} \def\ulb{{\underline b}} \def\ulc{{\underline c}}
\def\uld{{\underline d}} \def\ule{{\underline e}} \def\ulf{{\underline f}}
\def\ulg{{\underline g}} \def\ulm{{\underline m}}
\def\uln#1{\underline{#1}}
\def\ulp{{\underline p}} \def\ulq{{\underline q}} \def\ulr{{\underline r}}

\def\plpl{{+\!\!\!\!\!{\hskip 0.009in}{\raise -1.0pt\hbox{$_+$}}
{\hskip 0.0008in}}}

\def\mimi{{-\!\!\!\!\!{\hskip 0.009in}{\raise -1.0pt\hbox{$_-$}}
{\hskip 0.0008in}}}

\def\items#1{\\ \item{[#1]}}
\def\ul{\underline}
\def\un{\underline}
\def\-{{\hskip 1.5pt}\hbox{-}}

\def\kd#1#2{\d\du{#1}{#2}}
\def\fracmm#1#2{{{#1}\over{#2}}}
\def\footnotew#1{\footnote{\hsize=6.5in {#1}}}

\def\low#1{{\raise -3pt\hbox{${\hskip 1.0pt}\!_{#1}$}}}

\def\ip{{=\!\!\! \mid}}
\def\ze{\zeta^{+}}
\def\zeb{{\bar \zeta}^{+}}
\def\umb{{\underline {\bar m}}}
\def\unb{{\underline {\bar n}}}
\def\upb{{\underline {\bar p}}}
\def\um{{\underline m}}
\def\up{{\underline p}}
\def\Phib{{\Bar \Phi}}
\def\Phit{{\tilde \Phi}}
\def\Phibt{{\tilde {\Bar \Phi}}}
\def\Db{{\Bar D}_{+}}
\def\gg{{\hbox{\sc g}}}
\def\nt{$~N=2$~}

\border\headpic {\hbox to\hsize{March 1992 \hfill UMDEPP 92--187}}\par
\begin{center}
\vglue .25in

{\large\bf Extended ~Supersymmetry ~and ~Self--Duality \\
in ~2+2 ~Dimensions$\,$}\footnote{This work is supported in part
by NSF grant \# PHY-91-19746.} \\[.1in]

\baselineskip 10pt

\vskip 0.25in

S.~James GATES, Jr. ~and~ Hitoshi NISHINO \\[.1in]
{\it Department of Physics} \\ [.015in]
{\it University of Maryland at College Park}\\ [.015in]
{\it College Park, MD 20742-4111, USA} \\[.1in]
and\\[.1in]
{\it Department of Physics and Astronomy} \\[.015in]
{\it Howard University} \\[.015in]
{\it Washington, D.C. 20059, USA} \\[.18in]
and
\\[.18in]
Sergei V.~KETOV$\,$\footnote{On leave of absence from High Current Electronics
Institute of the Russian Academy of Sciences, Siberian Branch,
Akademichesky~4, Tomsk 634055, Russia.} \\[.1in]
{\it Department of Physics} \\[.015in]
{\it University of Maryland at College Park}\\[.015in]
{\it College Park, MD 20742-4111, USA} \\[.001in]

\vskip 1.0in

{\bf ABSTRACT}\\[.1in]
\end{center}

\begin{quotation}

{\baselineskip 5pt
The ~$N=2$~ supersymmetric {\it self-dual} Yang-Mills theory and the
{}~$N=4~$ and $~N=2$~ {\it self-dual} supergravities in ~$2+2$~ space-time
dimensions are formulated for the
first time. These formulations utilize solutions of the Bianchi identities
subject to the super-Yang-Mills or supergravity constraints in the relevant
$N$-extended superspace with the space-time signature $(2,2)$.
}

\endtitle
\def\doit#1#2{\ifcase#1\or#2\fi}
\def\[{\lfloor{\hskip 0.35pt}\!\!\!\lceil}
\def\]{\rfloor{\hskip 0.35pt}\!\!\!\rceil}
\def\delsl{{{\partial\!\!\! /}}}
\def\caldsl{{\calD\!\!\! /}}
\def\calO{{\cal O}}
\def\asym{({\scriptstyle 1\leftrightarrow \scriptstyle 2})}
\def\Lag{{\cal L}}
\def\du#1#2{_{#1}{}^{#2}}
\def\ud#1#2{^{#1}{}_{#2}}
\def\dud#1#2#3{_{#1}{}^{#2}{}_{#3}}
\def\udu#1#2#3{^{#1}{}_{#2}{}^{#3}}
\def\calD{{\cal D}}
\def\calM{{\cal M}}
\def\tildef{{\tilde f}}
\def\calDsl{{\calD\!\!\!\! /}}

\def\Hat#1{{#1}{\large\raise-0.02pt\hbox{$\!\hskip0.038in\!\!\!\hat{~}$}}}
\def\hati{{\hat{I}}}
\def\dt{$~D=10$~}
\def\alp{\alpha{\hskip 0.007in}'}
\def\oalp#1{\alp^{\hskip 0.007in {#1}}}
\def\naive{{{na${\scriptstyle 1}\!{\dot{}}\!{\dot{}}\,\,$ve}}}
\def\items#1{\vskip 0.05in\Item{[{#1}]}}
\def\item#1{\Item{#1}}

\def\pl#1#2#3{Phys.~Lett.~{\bf {#1}B} (19{#2}) #3}
\def\np#1#2#3{Nucl.~Phys.~{\bf B{#1}} (19{#2}) #3}
\def\prl#1#2#3{Phys.~Rev.~Lett.~{\bf #1} (19{#2}) #3}
\def\pr#1#2#3{Phys.~Rev.~{\bf D{#1}} (19{#2}) #3}
\def\cqg#1#2#3{Class.~and Quant.~Gr.~{\bf {#1}} (19{#2}) #3}
\def\cmp#1#2#3{Comm.~Math.~Phys.~{\bf {#1}} (19{#2}) #3}
\def\jmp#1#2#3{Jour.~Math.~Phys.~{\bf {#1}} (19{#2}) #3}
\def\ap#1#2#3{Ann.~of Phys.~{\bf {#1}} (19{#2}) #3}
\def\prep#1#2#3{Phys.~Rep.~{\bf {#1}C} (19{#2}) #3}
\def\ptp#1#2#3{Prog.~Theor.~Phys.~{\bf {#1}} (19{#2}) #3}
\def\ijmp#1#2#3{Int.~Jour.~Mod.~Phys.~{\bf {#1}} (19{#2}) #3}
\def\nc#1#2#3{Nuovo Cim.~{\bf {#1}} (19{#2}) #3}
\def\ibid#1#2#3{{\it ibid.}~{\bf {#1}} (19{#2}) #3}

\def\szet{{${\scriptstyle \b}$}}
\def\ula{{\un a}}
\def\ulb{{\un b}}
\def\ulc{{\un c}}
\def\uld{{\un d}}
\def\ulA{{\un A}}
\def\ulM{{\underline M}}
\def\cdm{{\Sc D}_{--}}
\def\cdp{{\Sc D}_{++}}
\def\vTheta{\check\Theta}
\def\Pisl{{\Pi\!\!\!\! /}}

\def\fracmm#1#2{{{#1}\over{#2}}}
\def\gg{{\hbox{\sc g}}}
\def\half{{\fracm12}}
\def\ha{\half}

\def\fracm#1#2{\hbox{\large{${\frac{{#1}}{{#2}}}$}}}
\def\Dot#1{{\hskip 0.3pt}{\raise 7pt\hbox{\large .}}
\!\!\!\!{\hskip 0.3pt}{#1}}

\def\Dot#1{\buildrel{_{_\bullet}}\over{#1}}
\def\dt#1{\Dot{#1}}

\oddsidemargin=0.03in
\evensidemargin=0.01in
\hsize=6.5in
\textwidth=6.5in

\noindent{\bf 1.~~Introduction}

The concept of {\it self-duality} (SD) in four space-time dimensions
happens
to be important in relation to exactly solvable models in lower dimensions.
Namely, the {\it self-dual Yang-Mills} (SDYM) theory [1] is likely to be
considered as the generating model for all integrable (bosonic) systems in
dimensions lower than four [2]. The SDYM theory and self-dual gravity turned
out to be closely related also with $N=2$ strings [3,4,5].
To consistently accomplish the SD, the relevant
space-time signature has to be $(2,2)$.

A natural task is to formulate the {\it supersymmetric} SD field theories in
 $2 + 2$ dimensions\footnotew{The $~2+2$~ dimensions in our notation
means four-dimensions with the signature $~(+,+,-,-)$, denoted also by
$~D=(2,2)$.} since the unification of
SD and supersymmetry
could result in more general solvable systems in lower dimensions and would
give rise to new generating theories encoding the underlying rich symmetry
structure of new integrable models. The key point allowing to construct the
SD field theories in a space-time with the $(2,2)$ signature is the existence
of the {\it Majorana-Weyl} (MW) spinors in such space-time, as we have noticed
recently [6]. The $N=1$ supersymmetric SDYM theory has been formulated, and the
 $N=1$ {\it self-dual supergravity} (SDSG) has been outlined [5,6].

A next question is about the
possibility of SDYM and SDSG with {\it extended} supersymmetries.
The purpose of this Letter is to develop the unification of SD
with {\it extended} supersymmetry by formulating the $N=2$ supersymmetric SDYM
theory and the $N=4$ SDSG in a space-time with the $(2,2)$ signature. We use
superspace methods to achieve this purpose.  Subsequently we will give the
$~N=2$~ SDSG.

\bigskip\bigskip

\noindent {\bf 2.~~$~N=2$~ SDYM}

We start with the
simplest example of SD with {\it extended} supersymmetry, namely
$~N=2$~ supersymmetric SDYM multiplet.

        Our $~N=2$~ YM has the field content $~(A\du \m
I,\l\du{\a i}I,\Tilde\l\du{\Dot\a i}I,S^I,T^I)$, where the indices
$~{\scst I,~J,~\cdots}$~ are for the
adjoint representation, and the indices $~{\scst i,~j,~\cdots~=~ 1,~2}$ are
for the two-dimensional representation of Sp(1) group.  The fields $~S^I$~
and $~T^I$~ are {\it real} scalars, while
$~\l\du{\a i}I$~ and $~\Tilde\l\du{\Dot\a i}I$~ are respectively chiral
and anti-chiral
{\it Majorana-Weyl} spinors, essentially in the same
notation\footnotew{For the vectorial indices of the local Lorentz frame
we use the {\it underlined} indices $~{\scst
\ula\,\equiv\,(a,~\Bar a),~\ulb\,\equiv\,(b,~\Bar b),~ \cdots}$, which
are equivalent to the pairs of 2-dimensional complex coordinate indices
$~{\scst a,~b,~\cdots= 1,~2}$~ and their complex
conjugates $~{\scst\Bar a,~\Bar b,~\cdots\,=\,\Bar1,~\Bar 2}$.
As for the spinorial indices, we use the {\it undotted} and {\it dotted}
indices $~{\scst\a,~\b,~\cdots\,=\, 1,~2~}$ and $~{\scst\Dot
\a,~\Dot\b,~\cdots\,=\,\Dot 1,~\Dot 2}$~ for Weyl spinors of each
chirality.} as the $~N=1$~
case [6].  We first solve superspace Bianchi identities (BIds)
{\it before} imposing any SD condition.
Our BIds are
$$\nabla_{\[A}F\du{B C)} I - T\du{\[A B|}D F\du {D|C)} I \equiv 0~~,
\eqno(2.1) $$
which are solved by the constraints
$$\eqalign{&F_{\a i\,\b j}{}^I = 2C_{\a\b} \,\e{\low{i j}}\, T^I ~~, ~~~~
F_{\Dot\a i\,\Dot\b j} {}^I = 2C_{\Dot\a\Dot\b} \,\e{\low {i j}}\, S^I ~~, \cr
& F_{\a i\,\ulb} = - i(\s_\ulb)_{\a\Dot\b} \Tilde\l\udu{\Dot\b} i I ~~,
{}~~~~F_{\Dot\a i\, \ulb} {}^I = - i(\s_\ulb)_{\b\Dot\a} \l\udu \b i I ~~, \cr
&F_{\a i\,\Dot\b j}{}^I = 0 ~~, ~~~~ T\du{\a i\,\Dot\b j} \ulc = i(\s^\ulc)
_{\a\Dot\b} \,\e{\low{i j}} ~~, \cr
& \nabla_{\a i} S^I = -\l\du{\a i} I ~~, ~~~~\Tilde\nabla_{\Dot\a i} S^I
= 0~~, \cr
&\Tilde\nabla_{\Dot\a i} T^I = -\Tilde\l _{\Dot\a i} {}^I ~~, ~~~~\nabla_{\a
i} T^I = 0~~, \cr
&\nabla_{\a i} \Tilde\l\du{\Dot\b j }I = - i \,\e{\low{i j}}\, (\s^\ulc)
_{\a\Dot\b}\nabla_\ulc T^I ~~, \cr
& \Tilde\nabla_{\Dot\a i} \l\du{\b j} I = + i \e{\low{i j}}\, (\s^\ulc)
_{\b\Dot\a} \nabla_\ulc S^I ~~, \cr
&\nabla_{\a i} \l\du{\b j}I = -\fracm 14 \,\e{\low{i j}} \,
(\s^{\ulc\uld} )_{\a\b}
F\du{\ulc\uld} I + f^{I J K} C_{\a\b} \,\e{\low{i j}}\, S^J T^K ~~, \cr
&\Tilde\nabla_{\Dot\a i} \Tilde\l\du{\Dot\b j} I = +\fracm 14 \,\e{\low {i
j}}\, (\s^{\ulc\uld}) _{\Dot\a\Dot\b} F\du{\ulc\uld} I + f^{I J K}
C_{\Dot\a\Dot\b} \e{\low {i j}}\, S^J T^K ~~, \cr }
\eqno(2.2)$$
where $~\e_{i j} $~ is the invariant antisymmetric tensor for the Sp(1).

        By the use of BIds of dimensions $~3/2\le d\le 2$, we get the
superfield equations:
$$\eqalign{&i(\s^\ula)_{\a\Dot\b} \nabla_\ula \Tilde\l^{\Dot\b i \,I}
- 2f^{I J K} \l\du\a {i J} T^K=0 ~~, \cr
&i(\s^\ula)_{\b\Dot\a} \nabla_\ula \l^{\b i\, I } - 2f^{I J K}
\Tilde\l\du{\Dot\a} {i\,J} S^K = 0~~, \cr
& \nabla_\ulb F^{\ula\ulb\, I} - 2 f^{I J K} (S^J \nabla^\ula T^K + T^J
\nabla^\ula S^K) - 2i f^{I J K} (\l^{i\,I}\s^\ula \Tilde\l\du i K ) = 0~~,
\cr
&\Bo S^I + f^{I J K} (\l^{i\, J} \l\du i K) - 4f^{I K J} f^{J L M} S^K
S^L T^M = 0 ~~, \cr
&\Bo T^I - f^{I J K} (\Tilde\l^{i\, J} \Tilde\l\du i K) + 4f^{I K J}f^{J L
M} S^L T^K T^M = 0 ~~. \cr}
\eqno(2.3) $$
An invariant lagrangian is given by
$$\eqalign{\Lag_{\rm SYM}^{N=2} = \,&-\fracm 14 (F\du{\ula\ulb}I)^2 - 2i
\Tilde\l^{i\,I}\s^\ulc \nabla_\ulc \l\du i I - 2(\nabla_\ula
T^I)(\nabla^\ula S^I) \cr
&+2f^{I J K} (\l^{i\,I} \l\du i J ) T^K - 2f^{I J K} (\Tilde \l^{I\, i}
\Tilde \l\du i I) S^K \cr
& + 4f^{I J K} f^{K L M} S^I S^L T^J T^M ~~. \cr }
\eqno(2.4) $$

        The system above is described {\it before} imposing our SD condition.
A word or two is in order regarding the Lagrangian in (2.4.). It can be
seen that the $S$-$T$ kinetic energy term is of a form that would ordinarily
imply the existence of ghosts (i.e. a breakdown in unitarity).  However,
since (2,2) spacetime already has {\it two} time-like coordinates, the
interpretation of unitarity in such theories is clearly outside of the
usual considerations.  As has been already mentioned in our previous paper
[6], our supersymmetric SD condition is to be
$$ S^I=0~~,  \eqno(2.5) $$
which triggers other equations like
$$w\du{\a i} I = 0~~, ~~~~ F_{\ula\ulb} = \half \e\du{\ula\ulb}{\ulc\uld}
F_{\ulc\uld} ~~,
\eqno(2.6) $$
the latter of which is the SD condition on $~F_{\ula\ulb}$~
equivalent to the three equations $~F_{a b} {}^I =
F_{\Bar a\Bar b}{}^I = 0~,~~\eta^{a\Bar b}F_{a\Bar b} {}^I=0$~ in the complex
coordinate notation [6].

        As in the $~N=1$~ case, we do {\it not} have any invariant
lagrangian in the usual sense for the self-dual system.  It may well be,
however, that there exists a nice superstring formulation related to
this $~N=2$~ system as its background, and only its path-integral, or
BRST covariant string field theory action gives
the action principle for such SDYM system, like the $~N=1$~ case [5].

\bigskip\bigskip

\noindent {\bf 3.~~$N=4$~ SDSG}

Since we have already seen how the ~$N=2$~ {\it extended} supersymmetric
SDYM system works, it
is straightforward to apply the same technique to the {\it extended
supergravity}.  Even though the simplest example is the $~N=2$~ SG,
since this
multiplet is easily obtained by a truncation from an $~N=4$~
supergravity, we give first the $~N=4$~ SG.

        Our on-shell $~N=4$~ extended supergravity with the {\it global}
SO(4) symmetry {\it before} the SD condition consists of the
fields $~(e\du \ula\ulm,\psi\du\ulm{\a i},\Tilde\psi\du\ulm{\Dot\a i},
A\du\ulm{i j},\L_{\a i},\Tilde\L_{\Dot\a i}, A,B)$, where the
gravitini $~\psi\du\ulm{\a
i}, ~\Tilde \psi\du\ulm{\Dot\a i}$~ and the spin 1/2 fields $~\L_i$,
$~\Tilde\L_i$~ are in the $~{\bf 4}\-$representation of the {\it global} SO(4)
group, while the central charge gauge fields $~A\du\ulm {i j}$~ are in the
$~{\bf 6}\-$representation.  Due to the similarity between the $~D=(1,3),
\,N=4$~ and $~D=(2,2),\,N=4$~ superspace, we can utilize the results in
Ref.~[7].  The essential difference, however, is that the superfields $~W$~
and $~\Bar W$~ in the former are now replaced by the two completely
independent real scalar superfields $~B$~ and $~A$, respectively.  This
sort of replacement seems to be rather universal, as long as the supersymmetry
in the system is {\it not} maximal one.  One other relevant fact, is that in
our construction of the $N=4$~ SDSG theory, it is
important to choose the ``correct'' version of $~D=(1,3),\,N=4$~ supergravity!
Among the three different superspace versions (the SO(4) theory, SU(4) theory
and the heterotic string related SU(4) theory [7]) it is {\it {only}} the
SO(4) theory that may be consistently truncated down to a self-dual version
of extended supergravity.

        The independent BIds are
$$\eqalign{&\nabla_{\[ A} T\du{B C)} D - T\du{\[ A B|}E T\du{E|C)} D
-R\du{\[ A B C)} D \equiv 0~~, \cr
& \nabla_{\[ A} F\du{B C)} {i j} - T\du{\[ A B|} D F\du{D|C)} {i j}
\equiv 0 ~~, \cr}
\eqno(3.1) $$
where $~F\du{A B}{i j}$~ are the superfield strengths for the the central
charge gauge fields.  These BIds are to be satisfied by some constraints, as
in the SDYM case.  Utilizing the similarities between the $~D=(1,3),\,N=4$~
and $~D=(2,2),\,N=4$, we can easily solve these BIds.  In fact, we see that
the following supertorsion, supercurvature, and constituent constraints can
solve the above BIds, where the supertorsion constraints  are
$$\eqalign{&T_{\uln{\a}\uln{\b}}{} ^{\uln{\g}} = \fracm 1 4 A
\L_{(\uln{\a}}\kd {\uln\b )} {\uln{\g}} ~~,~~~~
T_{\uln{\a}\uln{\b}}{} ^{\uln{\dt\g}} ~=~ C_{\a\b}C_{ijkl}
\Tilde\L^{\dt\g l} ~~,~~
T_{\uln{\a}\uln{\b}}{} ^{\uln{c}} ~=~ T_{\uln{\a}\uln{b}}{} ^{\uln{c}}
{}~=~ 0 ~,\cr
&T_{\uln{\a}\uln{\dt\b}}{} ^{\uln{\dt\g}} =
-~ \fracm 1 4 \d_{\uln{\dt\a}}{}^{\uln{\dt\g}} A \L_{\uln{\a}} ~~,~~
T_{\uln{\a}\uln{\dt\b}}{} ^{\ulc} ~=~ 2i \kd {\a} {\g}
\kd {\dt\a} {\dt\g} \kd i k ~~,~~
T_{\uln{a}\uln{\b}}{} ^{\uln{\dt\g}} ~=~ -~ iC_{\a\b}
\Tilde f_{\dt\a}{}^{\dt\g}{}_{jk} ~, \cr
&T_{\uln{a}\uln{\b}\uln{\g}} =
\fracm 14 \lrad{\a\Dot\a} C_{\b\g} \kd j k ~+~ i\half
C_{\a\b}\L_{\g l}\Tilde\L_{\dt\a}{}^{\[ k}
\d_j{}^{l \]} ~+~i \fracm 1 {16} C_{\a(\b}\L_{\g) l}
\Tilde\L_{\dt\a}{}^{l}\d_j{}^{k} ~~, \cr
&T_{\uln{a} \uln{b}}{} ^{\uln{\g}}  =  - C_{\dt\a \dt\b}
\bigg{[}\S_{\a \b}{}^{\g k} ~+~ i \fracm 13 a^{-1} \d_{(\a}{}^{\g}
(D_{\b)\dt\d} B )\Tilde\L^{\dt\d}{}^{k}\bigg{]}  + \fracm 1 4
C_{\a \b}\bigg{[}C^{klmn}\L^{\g}{}_{l}\Tilde f_{\dt\a \dt\b m n} ~+~
ia^{-1} (D^{\g}{}_{(\dt\a}B )\Tilde \L_{\dt\b) }{}^{k}\bigg{]} ~, \cr
&T_{\uln{a}\uln{b}}{} ^{\uln{c}}  =  ~-~i \fracm 1 2 \left[
\d_{\b}{}^{\g}\d_{\dt\a}{}^{\dt\g}\Tilde \L_{\dt\b}{}^{k}\L_{\a k} ~-~
\d_{\a}{}^{\g}\d_{\dt\b}{}^{\dt\g}\Tilde \L_{\dt\a}{}^{k}\L_{\b k}\right] ~.
\cr }
\eqno(3.2) $$
Here $~(A{\buildrel\leftrightarrow\over D}_{\a\Dot\a}B)
\equiv A(D_{\a\Dot\a} B) - (D_{\a\Dot\a} A) B~,~~a=a(A,B)\equiv
1-A^2-B^2$,
and we use the {\it underlined} indices $~{\scst \ul\a,~\ul\b,~\cdots}$~
and $~{\scst \ul{\Dot\a},~\ul{\Dot\b},~\cdots}$~ for $~{\scst \Dot\a
i,~\Dot\b i,~\cdots}$, which are temporarily different from the notation
in Ref.~[6].  Notice that there exist also complementary
constraints,\footnotew{Our $~A$~ and $~B$~ superfields correspond
respectively to $~\Bar W$~ and $~W$~ in Ref.~[7].  However, since
$~A$~ and $~B$~ are {\it not} related to each other by complex conjugation
in our $~D=(2,2)$, those complementary constraints are separately
needed.  The same is also true for the {\it dotted}$\leftrightarrow${\it
undotted} indices, as well as {\it tilded}$\,\leftrightarrow\,${\it
untilded} superfields.} obtained by exchanging
$~A\leftrightarrow B, ~\hbox{\it
dotted}\leftrightarrow \hbox{{\it undotted} indices}$, and {\it
tilded}$\leftrightarrow$\hbox{\it untilded} superfields, e.g.,
$~T\du{\ula\Dot\b}\g= - iC_{\Dot\a\Dot\b} f\dud\a\g{j k}$.  We omit
them in this Letter to save space.

The supercurvatures are:
$$\eqalign{&R_{\uln{\a} \uln{\b} \g\d}  =  0 ~~,~~~~
R_{\uln{\a} \uln{\b}\, \dt\g \dt\d}
{}~=~ 2C_{\a\b} \Tilde f_{\dt\g \dt\d i j} ~,~  \cr
&R_{\uln\a \uln{\dt\b} \g\d}
 =  - \fracm 1 2 C_{\a ( \g} \bigg{[} \L_{\d ) i}
\Tilde \L_{\dt\b}{}^{j}~-~\fracm 1 4 \kd i j \L_{\d) l}
\Tilde \L_{\dt\b}{}^{l}\bigg{]} ~, \cr
&R_{\uln\a , \b \dt\b, \g\d}  =  i \fracm {7} {16} C_{\a\b} C_{i j k l}
\Tilde \L_{\dt\b}{}^{j} f_{\g\d}{}^{kl} ~+~ i \fracm 1 {32} C_{\b ( \g|}
C_{i j k l} \Tilde \L_{\dt\b}{}^{j} f_{|\d )\a}{}^{kl} \cr
&~~~~~ ~~~~~ ~~~~~ +~ \fracm 3 {16} a^{-1} C_{( \g\vert ( \a}
(D_{\b)  \dt\b} A) \L_{|\d) i}
 ~-~ \fracm 5 {16} C_{\a\b}(D_{( \g\vert \dt\b} A)
\L_{\vert\d ) i}~, \cr
&~R_{\uln\a , \b\dt\b, \dt\g \dt\d}  =  -~ iC_{\a\b}
\Tilde \S_{\dt\b \dt\g \dt\d i} ~-~ i \fracm 1 {16} C_{\dt\b ( \dt\g}
C_{i j k l} \Tilde \L_{\dt\d )}{}^{j} f_{\a\b}{}^{kl} \cr
&~~~~~ ~~~~~ ~~~~~ +~ \fracm 1 {48} a^{-1} C_{\a\b} C_{\dt\b  ( \dt\g\vert}
(D_{\e \vert \dt\d)} A )
\L^{\e}{}_{i} ~+~ \fracm 3 {16} a^{-1} C_{\dt\b ( \dt\g\vert}
(D_{\a\vert \dt\d )} A )
\L_{\b ) i} ~, \cr} $$

\newpage
$$\eqalign{&R_{\a\dt\a , \b\dt\b, \g\d}  =  - \fracm 1 2 C_{\dt\a \dt\b}
\bigg{[} V_{\a\b\g\d} ~+~ i \fracm 1 {16} C_{ ( \g\vert ( \a}
\Tilde \L^{\dt\e l} \left\{ D_{\b ) \dt\e} ~+~ \fracm 3 4 a^{-1}
\lrad{\b )\dt\e} \right\} \L_{\vert\d ) l} \cr
&~~~~~ ~~~~~ ~~~~~ +~ C_{\a {(} \g} C_{\d ) \b} \bigg\{ \fracm 1 6 a^{-2} \big(
D_{\e \dt\z}  A \big{)(} D^{\e \dt\z} B  \big)
+ i \fracm 3{16} a^{-1} \lrad{\e\Dot \zeta} \L^{\Dot\zeta i} \L\ud\e i \cr
& ~~~~~ ~~~~~ ~~~~~ ~~~~~ ~~~~~ ~~~~~ ~+~ \fracm {15} {128}
\Tilde \L^{\dt\e l} \Tilde \L_{\dt\e}{}^{m} \L^{\z}{}_{l} \L_{\z m}
\bigg\} \bigg] \cr
& ~~~~~ ~~~~~ ~~~~~ ~+~ \fracm 1 2 C_{\a \b} \bigg{[}f_{\a\b i j}
\Tilde  f_{\dt\a \dt\b}{}^{i j} ~-~ \fracm 1 2 a^{-1} C_{\a \b}
\big{(}D_{(\g \vert (\dt\a\vert} A \big{)(}
D_{\vert\d ) \vert \dt\b ) } B \big{)} \cr
&~~~~~ ~~~~~ ~~~~~ ~~~~~ ~~~~~ ~~~ ~-~ i \fracm 1 {16} D_{(\g\vert (\dt\a\vert}
\left( \L_{\vert\d ) l} \Tilde \L_{\vert\dt\b )}{}^{l} \right)
{}~-~ \fracm {1} {64} \Tilde \L_{( \dt\a}{}^{l}
\Tilde \L_{\dt\b )}{}^{m}
\L_{\g l} \L_{\d m} \bigg{]} ~. \cr }
\eqno(3.3) $$

The central charge field strength are:
$$\eqalign{&F_{\ul\a\ul\b}{}^{k l} = 2a^{-1}C_{\a\b} E\du{i j}{k l} ~~, \cr
&E\du{i j}{k l} \equiv\half \left[ \d\du i{\[ k}\d\du j {l\]}
+ AC\du{i j}{k l} \right]
{}~~, ~~~~F\du{\a\Dot\b}{i j} = 0~~,  \cr
&F\du{\a\Dot\a\,\ul\b} {k l} = - i \half a^{-1/2} C_{\a\b}
\Tilde \L\du{\Dot\a} i C_{i j m n} \Tilde E^{m n k l} ~~, \cr
&F\du{\a\Dot\a\,\b\Dot\b} {k l} = \half a^{-1/2} \left[\, C_{\Dot\a\Dot\b}
f\du{\a\b}{m n} E\du{m n}{k l} + C_{\a\b}\Tilde  f_{\Dot\a\Dot\b\, m n}
\Tilde E^{m n k l} \,\right] ~~, \cr
& f\du{\a\b}{k l} = f\du{\b\a} {k l} ~~, \cr}
\eqno(3.4) $$
where $~f\du{\a\b}{i j}$~ and $~f\du{\Dot\a\Dot\b}{i j}$~ correspond to
the component YM field strength as their $~\theta=0\-$sectors.

The constituency relations  are:
$$\eqalign{&D_{\a i} B  =  (1-A^2 - B^2 ) \L_{\uln\a} ~~~,~~~
\Tilde D_{\Dot\a i} B = 0~~, \cr
&D_{\a i} \L_{\b j}  =  C_{i j k l} f_{\a\b}{}^{kl} ~-~ \fracm 3 4 A
\L_{\a i}\L_{\b j} ~~, ~~~~
{\Tilde D}_{\dt\a}{}^{i} \L_{\b j} ~=~ 2ia^{-1} \d_j{}^{i}(D_{\b
\dt\a} B ) ~+~ \fracm 3 4  B \Tilde\L_{\dt
\a}{}^{i} \L_{\b j} ~,\cr
&{\Tilde D}_{\dt\a}{}^{i}f_{\b\g}{}^{jk}  = \fracm 1 2 B
{\Tilde \L}_{\dt\a}{}^{i}f_{\b\g}{}^{jk} \cr
& ~~~~~ ~~~~~ ~~ ~+~ \fracm 1 2 C^{j k m n}\bigg{[}
\Tilde\L_{\dt\a}{}^{i} \L_{\b m} \L_{\g n}
{}~+~ i\d_m{}^{i}\left\{ D_{( \b\vert \dt\a}
 ~+~i \fracm 1{16} \Tilde\L_{\dt\a}{}^{l} \L_{(\b\vert l}
{}~+~ \fracm 3 4 a^{-1} \lrad{(\b|\Dot\a} \right\}\L_{\vert\g) n}
\bigg{]} ~, \cr
&D_{\a i}f_{\b\g}{}^{jk}  =  \d_i{}^{\[j} \S_{\a\b\g}{}^{k\]}
{}~-~ \fracm 1 2 A \L_{\a i} f_{\b\g}{}^{jk} ~+~ i \fracm 1 3a^{-1} \kd {i}
{\[j}
C_{\a (\b} (D_{\g) \dt\d} B )
\Tilde \L^{\dt\d k \]} ~, \cr
&{\Tilde D}_{\dt\a}{}^{i} \S_{\b\g\d}{}^{j}  =  \fracm 1 4 B
\Tilde \L_{\dt\a}{}^{i}
\S_{\b\g\d}{}^{j} ~-~ \fracm 1 6 \Tilde \L_{\dt\a}{}^{( i} \L_{
( \b |k} f_{|\g\d)}{}^{j)k} ~-~ i \fracm 1 6 C^{i j k l}
(D_{(\b\vert\dt\a} A )
\L_{\vert\g\vert k}\L_{\vert \d) l} \cr
&~~~~~ ~~~~~ ~~~~~ +~ i\bigg{[}D_{(\b\vert\dt\a}  ~-~
i \fracm 3 8 \Tilde \L_{\dt\a}{}^{l} \L_{( \b | l} ~+~ \half
a^{-1} \lrad{(\b|\Dot\a} \bigg{]} f_{\vert \g\d)}{}^{ij}
{}~,\cr
&D_{\a i}\S_{\b\g\d}{}^{j}  =  \d_{i}{}^{j} V_{\a\b\g\d} ~-~ \fracm 1 4 A
\L_{\a i} \S_{\b\g\d}{}^{j} ~-~ \fracm 1 6 a^{-2}
(D_{( \b \vert \dt\g} A) (D_{\vert \g \vert}{}^{\dt\g} B )
C_{\vert \d ) \a} \kd i j \cr
&~~~~~ ~~~~~ ~~~~~ +~ i \fracm 1 6 \bigg{[}\left\{ D_{(\b\vert \dt\g}
{}~+~ \fracm 34 a^{-1} \lrad{(\b|\Dot\g}
 ~+~i \fracm 1 {16} \Tilde \L_{\dt\g}{}^{l}
\L_{(\b\vert l} \right\} \L_{\vert\g\vert i}\bigg{]}\Tilde \L^{\dt\g j}
C_{\vert\d) \a} \cr
&~~~~~ ~~~~~ ~~~~~ -~ i \fracm 1 8 \bigg{[} \left\{ D_{(\b\vert \dt\g}
 ~+~ \fracm 34 a^{-1} \lrad{(\b\vert \dt\g}   ~+~ i\fracm 1 {16}
\Tilde \L_{\dt\g}{}^{l}
\L_{(\b\vert l} \right\} \L_{\vert\g\vert i}\bigg{]}\Tilde \L^{\dt\g k}
C_{\vert\d) \a}\d_{i}{}^j ~, \cr } $$

\newpage
$$\eqalign{&{\Tilde D}_{\dt\a}{}^{i} V_{\b\g\d\e}  =  \fracm 1 {24}
\bigg{[}2i \left\{ D_{(\b\vert\dt\a}  ~-~ \fracm 14
a^{-1}\lrad {(\b\vert\dt\a} \right\}
\S_{\vert \g\d\e )}{}^{i} ~+~ (\fracm 3 2 \Tilde\L_{\dt\a}{}^{i}
\L_{(\b| j} ~-~ \fracm {19} {16}\Tilde\L\du{\Dot\a} l \L_{(\b| l}\kd i j)
\S_{\vert \g\d\e )}{}^{j} \cr
&~~~~~ ~~~~~ ~~~~~ ~~~~~ ~~~~~ -~ \left\{ \fracm 9 {16} C_{k l m n}
\Tilde\L_{\dt\a}{}^{l}
f_{(\b\g|}{}^{mn} ~+~ i \fracm 12 a^{-1} \left( D_{(\b\vert\dt\a} A\right)
\L_{\vert\g\vert k} \right\}
f_{\vert\d\e)}{}^{ki} \bigg{]} ~, \cr
&D_{\a i} V_{\b\g\d\e}  =  - \fracm 1 {24} C_{\a (\b|} \bigg[
\left\{ D_{|\g |\dt\e} ~+~ \fracm 1 4a^{-1} \lrad{|\g |\dt\e} \right\}
\left\{ i \fracm 7 {16} C_{i k l m} \Tilde \L^{\dt\e k}f_{|\d\e)}{}^{lm}
{}~-~ a^{-1} (D_{|\d}{}^{\dt\e} A ) \L_{\e) i} \right\} \cr
& ~~~~~ ~~~~~ ~~~~~ ~~~~~ ~~~~~ -~ \fracm {15} {16} \left\{
\fracm 3 {16} C_{i k l m}
\Tilde \L^{\dt\e k}f_{|\g\d|}{}^{lm} ~+~ ia^{-1} (D_{|\g}{}^{\dt\e} A
\big{)} \L_{\d| i}\right\} \L_{|\e) n} \Tilde \L_{\dt\e}{}^n \bigg] ~. \cr}
\eqno(3.5) $$

        We are now ready to establish our SD condition, which yields the
SDSG with $~N=4$~ extended supersymmetry.  The principle is essentially
the same as the $~N=2$~ SDYM case (2.5), namely we can put one of the scalar
superfields to be zero, which is in our present notation:
$$B=0 ~~,
\eqno(3.6) $$
which triggers the other superfield equations, such as
$$\li{&\L_{\a i} = 0~~,
&(3.7{\rm a}) \cr
&f\du{\a\b}{i j} = 0~~,
&(3.7{\rm b}) \cr
&\S\du{\a\b}{\g k}= 0  ~~,
&(3.7{\rm c}) \cr
&V_{\a\b\g\d} = 0~~.
&(3.7{\rm d}) \cr } $$
Eq.~(3.7b) is equivalent to the SD condition on the central charge field
strength, (3.7c) is the $~N=4~$ analog of the SD gravitino field
strength condition $~W_{\a\b\g} = 0$~ of the $~N=1$~ case [6].  Finally
(3.7d) is equivalent to the SD condition of the Riemann tensor
$~R\du{\ula\ulb}{\ulc\uld} = (1/2) \e\du{\ula\ulb} {\ule\ulf}
R\du{\ule\ulf}{\ulc\uld}$.  This is further
equivalent to the Ricci-flatness $~R_{\ula\ulb} =
0$~ due to the vanishing torsion $~T\du{\ula\ulb}\ulc = 0$~
under (3.7a).
The consistency of the SD condition (3.6) with all the Bianchi identities can
be easily realized by the inspection of the above constraints, and in
particular the constituency relations (3.5).

\bigskip\bigskip

\noindent {\bf 4.~~$~N=2$~ SDSG}

        Once $~N=4$~ (SD)SG has been established, it is easy to obtain
it down to $~N=2$~ (SD)SG by truncation of appropriate superfields.  We
first perform the truncation {\it without} imposing the SD condition.
The resultant field content for the $~N=2$~ SG is
to be $~(e\du\ula\ulm,\psi\du\ulm{\a i},
\Tilde\psi\du\ula{\Dot\a i}, A\du\ulm{i j})$.
The main prescription here is to put the superfields $~\L_{\a
i},\,\Tilde\L_{\Dot\a i},\,A$~ and $~B~$ to zero.  Relevantly the
indices $~{\scst i,~j,~\cdots}$~ are now $~{\scst i,~j,~\cdots~=~1,~2}$~
for the SO(2) subgroup of the SO(4).

        After applying this prescription, and performing appropriate field
redefinitions, we get the superspace constraints:
$$\eqalign{&T\du{\a i\,\Dot\b j}\ulc = i \d_{i j}\, (\s^\ulc)_{\a\Dot\b}
{}~~, ~~~~ T_{\a i\,\b j}{}^\ulc = 0 ~~, \cr
&F_{\a i \,\b j} = \e{\low {i j}} C_{\a\b} ~~, ~~~~ F_{\Dot\a i \,\Dot\b j} =
\e{\low {i j}} C_{\Dot\a \Dot\b} ~~, ~~~~ F_{\a i\Dot\b j} = 0 ~~, \cr
&T\du{\a i\,\ulb} {\Dot\g k} = i\fracm 12 \e{\low i}{}^k
(\s_\ulc)\du\a{\Dot \g}
(F_{\ulb\ulc} + {\breve F}_{\ulb\ulc}) ~~, ~~~~ T\du{\a i \,\ulb}{\g k} =
0~~, \cr
&T\du{\Dot\a i \,\ulb} {\g k} = i\fracm 12 \e{\low i}{}^k (\s_\ulc) \ud
\g{\Dot\a} (F_{\ulb\ulc} - {\breve F}_{\ulb\ulc} ) ~~, ~~~~T\du{\Dot\a
i\, \ulb} {\Dot\g k} =0~~, \cr
&T\du{\ul\a\ul\b}{\ul\g} = 0~~, ~~~~ F_{\ul\a \ulb} = 0~~, ~~~~
T\du{\ula\ulb}\ulc =0~~, \cr
& R_{\a i\,\b j \, \ulc\uld} =
- \e{\low{i j}} \, C_{\a\b} (F_{\ulc\uld} +
{\breve F}_{\ulc\uld} ) ~~, ~~~~
R_{\Dot\a i\,\Dot\b j\, \ulc\uld} = -\e{\low{i j}}\, C_{\Dot\a\Dot\b}
(F_{\ulc\uld} - {\breve F}_{\ulc\uld} ) ~~, \cr
&R\du{\a i\,\b j\, \Dot\g k}{\Dot\d l} = \half C_{\a\b}
(\s^{\ule\ulf})\du{\Dot\g}{\Dot\d} \e_{i j} \d\du k l F_{\ule\ulf} ~~, ~~~~
R\du{\Dot\a i\, \Dot\b j\, \g k} {\d l} = \half C_{\Dot\a\Dot\b}
(\s^{\ule\ulf}) \du\g\d \e_{i j} \d\du k l F_{\ule\ulf} ~~, \cr
&R\du{\a i\,\b j\, \g k}{\d l} = 0~~, ~~~~ R\du{\a i \,\Dot\b j \,
\ul\g}{\ul\d} =0~~. }
\eqno(4.1) $$
where we are using the {\it underlined} spinorial indices $~{\scst \ul\a,
{}~\ul\b,~\cdots}$~ for the abbreviation of the pairs of indices
$~{\scst (\a i,~\Dot\a i),~(\b i,~\Dot\b i),~\ldots}\,$.  The graviphoton
field strength $~F_{\ula\ulb}$~ is equivalent to $~f\du{\ula\ulb}{1 2}$~ in
the $~N=4$~ notation in section 3.  The $~{\breve F}_{\ula\ulb} \equiv
(1/2)\e\du{\ula\ulb}{\ulc\uld} F_{\ulc\uld}$~ is the {\it dual}~ of
$~F_{\ula\ulb}$.

        It is convenient to use the {\it chiral} superfield $~W_{\a\b}$~ and
the {\it anti-chiral} superfield $~\Tilde{W}_{\Dot\a\Dot\b}$~ defined by
$$W_{\a\b} \equiv \half (\s^{\ulc\uld})_{\a\b} F_{\ulc\uld}~~, ~~~~
\Tilde W_{\Dot\a\Dot\b} \equiv \half (\s^{\ulc\uld})_{\Dot\a\Dot\b}
F_{\ulc\uld}~~,
\eqno(4.2)$$
as the analogues of the $W_{\a\b\g}$~ and $~\Tilde{W}_{\Dot\a\Dot{\b}\Dot{\g}}
$~ in the $~N=1$~ SG [6].

        It is straightforward to see that the superfield equations are
now
$$\li{&i(\s^{\ula\ulb\ulc}) _{\b\Dot\a} T\du{\ulb\ulc}{\b j} =0 ~~,
{}~~~~ i(\s^{\ula\ulb\ulc})_{\a\Dot\b} \Tilde T\du{\ulb\ulc}{\Dot\b j} =
0~~, &(4.3{\rm a}) \cr
&\nabla_\ula F^{\ula\ulb} = 0~~,
&(4.3{\rm b}) \cr
&R_{\ula\ulb} + \half \left(F_{\ula\ulc} F\du\ulb\ulc - {\breve F}
_{\ula\ulc} {\breve F} \du\ulb\ulc \right)  \cr
&~~~~~ = R_{\ula\ulb} + \left[\, F_{\ula\ulc} F\du \ulb\ulc - \fracm 14
\eta_{\ula\ulb} (F_{\ulc\uld}{})^2 \, \right] = 0~~,
&(4.3{\rm c}) \cr }
$$
This system is in fact quite similar to the previously considered cases of
$N=1$ YM and $N=1$ SG [5], as well as their $N=2$ and $N=4$ counterparts
above, making it easier to impose a SD condition.   Eq.~(4.3c)
emphasizes a salient point, i.e. when the SD condition is imposed on
any of our realizations of extended supersymmetry, the energy-momentum tensor
that appears in the Einstein equation of motions is seen to vanish. This same
statement can be seen to be true of the source current to the YM gauge field
in (2.3). It too vanishes in the presence of the supersymmetric SD condition.

Our SD condition for the $N=2$ SDSG reads
$$W_{\a\b} = 0~~.
\eqno(4.4) $$
Inspecting all the BIds, we see that this condition implies the equations
$$\eqalign{&T\du{\ula\ulb}{\g k} = 0~~, ~~~~ \Tilde T\du{\ula\ulb} {\Dot\g k}
=
\half \e\du{\ula\ulb} {\ulc\uld} \Tilde T\du{\ulc\uld} {\Dot\g k} ~~, \cr
&F_{\ula\ulb} = \breve F_{\ula\ulb} ~~, \cr
&R_{\ula\ulb} = 0~~, ~~~~R\du{\ula\ulb} {\ulc\uld} = \half \e
\du{\ula\ulb}{\ule\ulf} R\du{\ule\ulf}{\ulc\uld} ~~. \cr}
\eqno(4.5) $$

        There is no essential fundamental difference between the $~N=1$~ SDSG
of Ref.~[4], the $~N=2$~ SDSG just considered and the $~N=4$~ SDSG formulated
above as for the way they constructed, and everything is straightforward.
The actual construction of the $N=4$ SDSG does imply the existence of the
$N=2$ SDSG by an appropriate truncation of the former,
and it gives another straightforward way to get the latter.

\bigskip\bigskip

\noindent{\bf 5.~Concluding Remarks}

The explicit superspace construction of the $N=2$ supersymmetric SDYM and
the $N=2$ and $N=4$ SDSGs proves the consistency between SD and extended
supersymmetry in $2 + 2$ dimensions. The indefinite $(2,2)$ signature of
space-time and the existence of Majorana-Weyl spinors in this space-time were
crucial in all of the constructions.  The real difference between the
superspace formulations of various supersymmetric gauge theories and
supergravities with extended supersymmetry in $~D=(1,3)$~ and $~D=(2,2)$~
stems from the {\it reality conditions} imposed on the spinor
derivatives,
the superspace torsion and curvature components with spinor indices.  The case
of {\it maximally} extended supersymmetry or supergravity subject to the
SD condition needs a special treatment [8].

It is worthwhile to mention that, contrary to naive expectations, the SD
conditions on the YM field strength and the Riemann curvature are {\it not}
modified within extended supersymmetry, but just accompanied by the associated
equations on the other components of the self-dual supermultiplet under
consideration. This is particularly amusing in extended SDSGs, where both
scalars and vectors are added to the graviton field, and the SD gravity
condition {\it implies} the SDYM condition.

        Both SDYM and self-dual gravity equations of motions are the
non-linear field equations, which are known to be integrable.  Hence their
supersymmetric generalizations should be also integrable.  In fact,
four-dimensional self-dual theories can be identified with
{\it two-dimensional} non-linear sigma-models on twistor surfaces with
infinite dimensional gauge groups [9], and this link can in principle be
supersymmetrized.  The relevant gauge symmetries are just area-preserving
diffeomorphisms [10], which also indicates possible connection to the
$~W_{\infty}\-$algebra [11].  Our construction of $~N=4$~ SDSG theory gives
the first evidence for the existence of an $~N=4$~ $~W_\infty\-$algebra.

\vfill\eject

\refs

\normalsize

\items{1} A.A.~Belavin, A. M. Polyakov, A. Schwartz and Y. Tyupkin,
\pl{59}{75}{85};
\item{  } R.S.~Ward, \pl{61}{77}{81};
\item{  } M.F.~Atiyah and R.S.~Ward, \cmp{55}{77}{117};
\item{  } E.F..~Corrigan, D.B.~Fairlie, R.C.~Yates and P.~Goddard,
\cmp{58}{78}{223};
\item{  } E.~Witten, \prl{38}{77}{121};
\item{  } A.N.~Leznov and M.V.~Saveliev, \cmp{74}{80}{111};
\item{  } L.~Mason and G.~Sparling, \pl{137}{89}{29};
\item{  } I.~Bakas and D.A.~Depireux, \mpl{A6}{91}{399}; {\it ibid.} 1561;
2351.
\items{2} R.S.~Ward, Phil.~Trans.~Roy.~Lond.~{\bf A315} (1985) 451 ;
\item{  } N.J.~Hitchin, Proc.~Lond.~Math.~Soc.~{\bf 55} (1987) 59 ;
\item{  } A.A.~Belavin and V.E.~Zakharov, \pl{73}{78}{53}.

\items{3} H.~Ooguri and C.~Vafa, \mpl{A5}{90}{1389};
\np{361}{91}{469}; \ibid{\bf B367}{91}{83}.

\items{4} H.~Nishino and S.J.~Gates, Jr., Maryland preprint, UMDEPP
92-137 (January, 1992).

\items{5} H.~Nishino, S.J.~Gates, Jr.~and S.V.~Ketov, Maryland preprint,
UMDEPP 92--171 (February, 1992).

\items{6} S.V.~Ketov, H.~Nishino and S.J.~Gates, Jr., Maryland preprint,
UMDEPP 92--163 (February, 1992).

\items{7} S.J.~Gates, Jr., \np{218}{83}{409}; S.J.~Gates, Jr. and
J.~Durachta, Mod. Phys. Lett. {\bf {A21}} (1989) 2007.

\items{8} S.V.~Ketov, H.~Nishino and S.J.~Gates, Jr., Maryland preprint,
in preparation (April, 1992).

\items{9} Q.-H.~Park, \pl{238}{90}{287}; \ibid{257B}{91}{105}.

\items{10} Q.-H.~Park, Cambridge preprint, DAMTP R-91/12 (October, 1991).

\items{11} E.~Sezgin, Texas A \& M preprint, CTP-TAMU-13/92 (February,
1991).

\end{document}